\documentclass[journal,transmag]{IEEEtran}
\usepackage[T1]{fontenc}  
\usepackage[english]{babel}
\usepackage[backend=biber, style=ieee]{biblatex}
\usepackage{graphicx}
\usepackage{subcaption}
\usepackage{amsmath}
\usepackage{amssymb}
\usepackage{nicefrac}
\usepackage{xspace}
\usepackage{tikz}
\usepackage{pgfplots}
\usepackage{hyperref}

\usetikzlibrary{patterns}

\bibliography{./magnum_library.bib}
\AtBeginBibliography{\small}

\newcommand{\adjustedaccent}[1]{%
	\mathchoice{}{}
	{\mbox{\raisebox{-.5ex}[0pt][0pt]{$\scriptstyle#1$}}}
	{\mbox{\raisebox{-.35ex}[0pt][0pt]{$\scriptscriptstyle#1$}}}
}
\newcommand\bow[1]{\overset{\adjustedaccent{\smallfrown}}{#1}}

\newcommand{\dd}{\mathrm{d}}
\newcommand{\rr}{\vec{r}}
\newcommand{\Afield}[1]{\vec{A}_\mathrm{#1}}
\newcommand{\afield}[1]{\vec{A}_\mathrm{#1}}
\newcommand{\Hfield}[1]{\vec{H}_\mathrm{#1}}
\newcommand{\hfield}[1]{\vec{H}_\mathrm{#1}}
\newcommand{\Va}{V_\mathrm{a}}
\newcommand{\Vi}{V_\mathrm{i}}
\newcommand{\nvec}{\vec{n}}
\newcommand{\curl}{\mathrm{curl}}
\newcommand{\nxhm}{\nvec \times \Hfield{m}}
\newcommand{\abow}{\bow{a}}

\begin{document}
	\title{Efficient Reduced Magnetic Vector Potential Formulation for the Magnetic Field Simulation of Accelerator Magnets }%
	\author{\IEEEauthorblockN{Laura A.~M.~D'Angelo\IEEEauthorrefmark{1,2},
			Dominik Moll\IEEEauthorrefmark{1}, Andrea Vitrano\IEEEauthorrefmark{3}, Nicolas Marsic\IEEEauthorrefmark{4}, Erik Schnaubelt\IEEEauthorrefmark{1,2,3},\\ Mariusz Wozniak\IEEEauthorrefmark{3},
			Herbert De Gersem\IEEEauthorrefmark{1,2} and Bernhard Auchmann\IEEEauthorrefmark{5}}
		\IEEEauthorblockA{\IEEEauthorrefmark{1}Institute for Accelerator Science and Electromagnetic Fields, Technische Universität Darmstadt, Darmstadt, Germany}
		\IEEEauthorblockA{\IEEEauthorrefmark{2}Graduate School Computational Engineering, Technische Universität Darmstadt, Darmstadt, Germany}
		\IEEEauthorblockA{\IEEEauthorrefmark{3}CERN, Meyrin, Switzerland}
		\IEEEauthorblockA{\IEEEauthorrefmark{4}Dassault Systèmes, Darmstadt, Germany}
		\IEEEauthorblockA{\IEEEauthorrefmark{5}Paul Scherrer Institute, Villigen, Switzerland}
		\thanks{Manuscript received Month Day, Year; revised Month Day, Year. 
			Corresponding author: Laura A.~M.~D'Angelo (email: dangelo@temf.tu-darmstadt.de).}}
	
	\markboth{IEEE Transactions on Magnetics, Vol.~X, No.~Y, Month~Year}%
	{D'Angelo \MakeLowercase{\textit{et al.}}: Paper Title}
	
	\maketitle 
	
	\begin{abstract}
		The major advantage of reduced magnetic vector potential formulations (RMVPs) is that complicated coil structures do not need to be resolved by a computational mesh. Instead, they are modeled by thin wires, whose source field is included into the simulation model along Biot-Savart's law. Such an approach has already been successfully employed in ROXIE for the simulation of superconducting Large Hadron Collider magnets at CERN. This work presents an updated RMVP approach, which significantly outperforms the original method. The updated formulation is postulated, implemented, verified, compared to the original formulation, and applied for the simulation of a quadrupole magnet. The promising results of this work encourage further investigation towards an updated simulation framework for next-generation accelerator magnets.
	\end{abstract}	
	
	
	\section{Introduction} 
	\IEEEPARstart{H}{igh-temperature} superconducting (HTS) technology is expected to have a significant impact on next-generation synchrotrons~\cite{Durante_2018aa}. With this step forward, however, magnet design is confronted with new challenges regarding the design of large, high-field and high-quality magnet systems. 
	Computer-aided design and numerical field simulation generally play a crucial role in designing and optimizing accelerator magnet systems, and will do even more so regarding the next-generation HTS magnet systems. 
	For the past decades, the simulation software ROXIE~\cite{Russenschuck_1998aa} proved to be an indispensable workhorse for designing the low-temperature superconducting (LTS) magnets of the Large Hadron Collider (LHC). ROXIE combines a hybrid finite-element (FE) boundary-element method with a reduced magnetic vector potential (RMVP) formulation~\cite{Biro_1990aa, Paul_1997aa}, leading to fast and accurate simulations~\cite{Biro_1999ab, Kurz_2000aa}. 
	Herein, the coils are modeled as thin wires, and their excitation is included as a source magnetic field, which is calculated by Biot-Savart's law. The major advantage of this formulation is that these wires do not need to be resolved by a computational mesh. Especially superconducting accelerator magnets, which typically contain hundreds or thousands of coil windings, greatly benefit from this approach.
	
	Nonetheless, ROXIE and commercial out-of-the-box simulation tools struggle with the multi-scale nature that is imposed by HTS tapes, resulting into excessive computation times~\cite{Grilli_2016aa}. The goal of this work is to improve the computational efficiency of the RMVP approach and to contribute towards suitable simulation tools for future HTS magnet design campaigns. 
	To do so, ROXIE's original formulation is first adapted to a pure FE model in Sec.~\ref{sec:original_rmvp}, and then reformulated in order to diminish the number of Biot-Savart integrals to be calculated to quantify the source magnetic field. This leads to a multi-step calculation procedure, which is presented in Sec.~\ref{sec:rmvp} and demonstrated for a case study with an eccentric line current in an iron tube. In Sec.~\ref{sec:studies}, the updated RMVP formulation is analyzed regarding accuracy and performance. Herein, the proposed procedure proves to be clearly superior to ROXIE's original formulation. Sec.~\ref{sec:quadrupole} finally showcases the updated RMVP formulation by applying it to the two-dimensional (2D) nonlinear magnetostatic simulation of the LHC's LTS MQXA quadrupole magnet~\cite{Ostojic_2002aa}. All simulations have been carried out using the freely available FE solver GetDP~\cite{Dular_1998aa}.

	\section{Original RMVP Formulation}\label{sec:original_rmvp}
	In this section, the original RMVP formulation from~\cite{Biro_1990aa} is recapitulated. 
	The original physical problem that has to be solved is the magnetostatic problem with a homogeneous Dirichlet boundary condition, reading
	\begin{subequations}
		\begin{align}
			\nabla \times \left( \nu \nabla \times \Afield{} \right) &= \vec{J} & \text{in $V$}, \\
			\nvec \times \Afield{} &= 0 & \text{on $\partial V$}.
		\end{align}%
		\label{eq:volumetric}%
	\end{subequations}%
	Herein, $\Afield{}$ is the sought-for magnetic vector potential (MVP), $\nu$ is the reluctivity, which is possibly nonlinear, $\nu = \nu(\rr, \vec{B}(\rr))$, and $\vec{J}$ is the current density, which represents the excitation in this problem. The computational domain $V = \Va \cup \Vi$ consists of the coil and \underline{a}ir domain $\Va$ and the \underline{i}ron domain $\Vi$, and $\partial V$ is its boundary. Classically, the current excitation is modeled in the right-hand side $\vec{J}$, e.g.~by using winding functions~\cite{Schops_2013aa}. This procedure requires the explicit discretization of the individual wires (or at least half-turns) in the FE mesh. 
	
	In contrast, the RMVP method represents the wires by one-dimensional (1D) curves, which are not necessarily taken into account in the FE mesh. This benefits the meshing workload in the overall simulation process.
	The magnetic vector potential (MVP) is decomposed into
	\begin{equation}
		\Afield{} = \Afield{s} + \Afield{r},
		\label{eq:original_decomposition}
	\end{equation}
	where $\Afield{s}$ is called the \underline{s}ource MVP, and $\Afield{r}$ the \underline{r}educed MVP. $\Afield{s}$ is obtained by evaluating Biot-Savart's law~\cite{Jackson_1998aa} 
	\begin{equation}
		\Afield{s} = \frac{\mu_0}{4\pi} \int\limits_{\mathcal{L}'} \frac{I \, \dd \vec{s}'}{|\vec{r}-\vec{r}'|}
		\label{eq:biot-savart-integral}
	\end{equation}
	for \emph{all} spatial coordinates $\vec{r} \in V$. 
	The source domain $\mathcal{L}'$ represents the line, on which the line current $I$ is located. In a three-dimensional (3D) model, this would be an arbitrarily complicated 1D curve loop in space; while in a two-dimensional (2D) setting, $\mathcal{L}'$ is reduced to a zero-dimensional (0D) point, which represents a line current going in or out of plane. Furthermore, in 2D, the MVP is assumed to have only a $z$-component, $\Afield{} = A_z(x,y) \vec{e}_z$, and Biot-Savart's law~\eqref{eq:biot-savart-integral} becomes
	\begin{equation}
		A_z = \frac{\mu_0}{2\pi} \int\limits_{\mathcal{L}'} I \ln( \left| \rr - \rr' \right|^{-1} ) \, \dd r'.
	\end{equation}
	Multiple sources are taken into account by superposition. 
	
	Eventually, the discrete source MVP 
	\begin{equation}
		\Afield{s}(\rr) \approx \sum\limits_{j=1}^{N_\text{edge}} \abow_{\mathrm{s},j} \, \vec{w}_j(\rr)
	\end{equation}
	can be computed on the mesh edges $j=1,\dots,N_\text{edge}$ in two ways: One can utilize the partition-of-unity property and calculate the discrete coefficients $\abow_{\mathrm{s},j}$ per edge $e_j$ directly by weighting~\eqref{eq:biot-savart-integral} with the $j$-th edge function $\vec{w}_j$ and integrating over that edge $e_j$,
	\begin{equation}
		\abow_{\mathrm{s},j} = \frac{\mu_0}{4\pi} \int\limits_{e_j} \int\limits_{\mathcal{L}'} \frac{I \, \dd \vec{s}'}{|\vec{r}-\vec{r}'|} \cdot \vec{w}_j \, \dd s.
	\end{equation}
	Alternatively, one performs a weak $L^2$-projection of the Biot-Savart integral onto $\Afield{s}$,
	\begin{equation}
		(\Afield{s}, \Afield{s}')_V = \left( \frac{\mu_0}{4\pi} \int\limits_{\mathcal{L}'} \frac{I \, \dd \vec{s}'}{|\vec{r}-\vec{r}'|}, \Afield{s}' \right)_V
	\end{equation}
	with test functions $\Afield{s}' \in H(\text{curl};V)$ in the Hilbert space~\cite{Monk_2003aa} 
	\begin{equation}
		H(\text{curl};V) := \lbrace \Afield{} \in L^2(V) ~:~ \nabla \times \Afield{} \in L^2(V) \rbrace.
	\end{equation}
	This work uses the built-in $L^2$-projection of GetDP.
	
	The reduced MVP $\Afield{r}$ is computed by solving the boundary value problem (BVP)
	\begin{subequations}
		\begin{align}
			\nabla \times (\nu \nabla \times \Afield{r}) &= -\nabla \times (\nu \nabla \times \Afield{s}) & \text{in $\Vi$}, \\
			\nabla \times (\nu_0 \nabla \times \Afield{r}) &= 0 & \text{in $\Va$}, \\
			\vec{n} \times \Afield{r} &= -\vec{n} \times \Afield{s} & \text{on $\partial V$}. \label{eq:original_bc}%
		\end{align}%
		\label{eq:original_strong}
	\end{subequations}%
	Using a Ritz-Galerkin approach, the weak formulation is obtained as: Find $\Afield{r} \in H_\mathrm{r}(\text{curl};V)$, s.t.
	\begin{equation}
			(\nu \nabla \times \Afield{r}, \nabla \times \Afield{r}')_V = -(\nu \nabla \times \Afield{s}, \nabla \times \Afield{r}')_{\Vi}
			\label{eq:original_weak}
	\end{equation}
	$\forall \Afield{r}' \in H_\mathrm{r}(\text{curl};V)$, where $\Afield{r}'$ is a test function, and 
	\begin{equation}
			H_\mathrm{r}(\text{curl};V) = \lbrace \Afield{} \in H(\text{curl};V) : \gamma_{\partial V}(\Afield{}) = -\nvec \times \Afield{s} \rbrace
	\end{equation}
	is chosen in order to fulfill~\eqref{eq:original_bc}. Herein,  
	\begin{equation}
		\gamma_{\mathcal{B}}(\Afield{}) = \vec{n} \times \Afield{} |_{\mathcal{B}}
	\end{equation}
	is the tangential trace operator w.r.t.~a boundary $\mathcal{B}$~\cite{Monk_2003aa}.

	The weak formulation~\eqref{eq:original_weak} is solved by a FE method employing standard edge shape functions. Lastly, the total MVP $\Afield{}$ is composed of $\Afield{s}$ and $\Afield{r}$ following~\eqref{eq:original_decomposition}. Note that the Biot-Savart integral~\eqref{eq:biot-savart-integral} must be evaluated in the whole domain $V$, whether one is actually interested in the solution of the whole domain or of only a small sub-domain.

	\section{Updated RMVP Formulation}\label{sec:rmvp}
	\subsection{Idea and derivation}\label{subsec:derivation}
	The domain $V$ is decomposed into a non-permeable sub-domain $\Va$ (consisting of e.g.~\underline{a}ir) and a source-free sub-domain $\Vi$ (containing e.g.~\underline{i}ron). Thereby, $\Va$ is fully enclosed by $\Vi$, and $\Gamma$ denotes the interface between those domains. This configuration is illustrated for a simply 2D case in Fig.~\ref{fig:ecc_setup}. The BVP~\eqref{eq:original_strong} is expressed for both domains separately and, after introduction of the source MVP, reads
	\begin{subequations}
		\begin{align}
			\nabla \times \left( \nu_0 \nabla \times \left( \Afield{a} - \Afield{s} \right) \right) &= 0 & \text{in $\Va$}, \\
			\nabla \times \left( \nu \nabla \times \Afield{i} \right) &= 0 & \text{in $\Vi$}, \\
			\nvec \times \Afield{a} &= \nvec \times \Afield{i} & \text{at $\Gamma$}, \\
			\nvec \times \Hfield{a} &= \nvec \times \Hfield{i} & \text{at $\Gamma$}, \\
			\nvec \times \Afield{i} &= 0 & \text{at $\partial V$},
		\end{align}%
	\end{subequations}%
	where $\Afield{a}$ and $\Afield{i}$ are the MVPs and $\Hfield{a}$ and $\Hfield{i}$ the magnetic field strengths in $\Va$ and $\Vi$, respectively. Then, an additional field $\Afield{m}$ is added to the total MVP in $\Va$, which is the solution of a so-called image problem as will be described in Sec.~\ref{subsec:image}, such that homogeneous boundary conditions are enforced at $\Gamma$, yielding the BVP
	\begin{subequations}
		\begin{align}
			\nabla \times \left( \nu_0 \nabla \times \Afield{b} \right) &= 0 & \text{in $\Va$}, \\
			\nabla \times \left( \nu \nabla \times \Afield{i} \right) &= 0 & \text{in $\Vi$}, \\
			\nvec \times \Afield{b} &= \nvec \times \Afield{i} & \text{at $\Gamma$}, \\
			\nvec \times \Hfield{b} + \nvec \times \left( \Hfield{s} + \Hfield{m} \right) &= \nvec \times \Hfield{i} & \text{at $\Gamma$}, \\
			\nvec \times \Afield{i} &= 0 & \text{at $\partial V$},
		\end{align}%
		\label{eq:multi-domain-form}%
	\end{subequations}%
	with $\Afield{b} = \Afield{a} - \Afield{s} - \Afield{m}$ and $\Hfield{b} = \Hfield{a} - \Hfield{s} - \Hfield{m}$. Hence, the sub-domain solutions $\Afield{b}$ and $\Afield{i}$ are tangentially continuous at $\Gamma$. Their co-normal derivatives (magnetic field strengths), however, feature a jump prescribed by $\Hfield{s} + \Hfield{m}$. This jump can be interpreted as a surface current density in A/m, 
	\begin{equation}
		\vec{K}_\mathrm{g} = \nvec \times \left( \Hfield{s} + \Hfield{m} \right). 
		\label{eq:surface_current_density}
	\end{equation}
	By substituting $\Afield{g} = \Afield{b}$ in $\Va$ and $\Afield{g} = \Afield{i}$ in $\Vi$, the multi-domain formulation~\eqref{eq:multi-domain-form} can be transferred to a single-domain formulation w.r.t.~the so-called reaction MVP $\Afield{g}$, in which $\vec{K}_\mathrm{g}$ is imposed on the interface $\Gamma$. This so-called reaction sub-problem will be introduced in detail in Sec.~\ref{subsec:reaction}.
	
	This approach resembles the sub-domain FEM based on a perturbation technique, where a missing continuity at the interface of two sub-domains is also restored via a jump in the tangential magnetic field strength expressed as a surface current density source~\cite{Dular_2010aa}.

	\subsection{Ansatz and Biot-Savart sub-formulation}
	As a result of the previous section, the total MVP $\Afield{}$ is decomposed into
	\begin{equation}
		\Afield{} = \left\lbrace 
		\begin{array}{ll}
			\Afield{g} + \Afield{s} + \Afield{m} & \text{in $\Va$}, \\
			\Afield{g} & \text{in $\Vi$},
		\end{array}%
		\right. \label{eq:ansatz}
	\end{equation}%
	where $\Afield{s}$ is called the \underline{s}ource MVP, $\Afield{m}$ the i\underline{m}age MVP, and $\Afield{g}$ the reaction MVP. 
	$\Afield{s}$ is obtained by evaluating Biot-Savart's law~\eqref{eq:biot-savart-integral} for all spatial coordinates $\vec{r} \in \Va$ that are of interest, but at least for $\vec{r} \in \Gamma$. This is the major advantage to the original RMVP approach~\cite{Biro_1990aa}, where $\Afield{s}$ needed to be calculated in the whole domain $V$.
	\subsection{Image sub-formulation}\label{subsec:image}
	The image MVP $\Afield{m}$ is the solution of the BVP
	\begin{subequations}
		\begin{align}
			\nabla \times ( \nu_0 \nabla \times \Afield{m} ) &= 0 & \text{in $\Va$}, \\
			\nvec \times \Afield{m}+\nvec \times \Afield{s} &= 0 & \text{on $\Gamma$}. \label{eq:image_bc_strong}%
		\end{align}%
		\label{eq:image_subform_strong}%
	\end{subequations}%
	This sub-formulation originates from using the method of images~\cite{Jackson_1998aa}. In this context, $\Afield{m}$ represents the field that needs to be added to $\Afield{s}$ in $\Va$ in order to enforce electric boundary conditions at $\Gamma$.
	In other words, the MVP subtotal $\Afield{s} + \Afield{m}$ is the equivalent of a Green's function obeying a homogeneous Dirichlet boundary condition at $\Gamma$~\cite{Jackson_1998aa}.
	
	Using a Ritz-Galerkin approach and $\nxhm$ for the tangential component of the image magnetic field strength, the weak formulation is obtained as:\\ 
	Find $\afield{m} \in H(\curl;\Va), \nxhm \in H^{-1/2}(\curl; \Gamma)$, s.t. 
	\begin{subequations}
		\begin{align}
			( \nu_0 \nabla \times \afield{m}, \nabla \times \afield{m}')_{\Va} + (\nxhm, \afield{m}')_{\Gamma} &= 0, \label{eq:image_pde_weak} \\
			(\afield{m}, \nxhm')_{\Gamma} + (\afield{s}, \nxhm')_{\Gamma} &=0 \label{eq:image_bc_weak}
		\end{align}%
		\label{eq:image_subform_weak}%
	\end{subequations}%
	$\forall \afield{m}' \in H(\curl;\Va), \forall \nxhm' \in H^{-1/2}(\curl;\Gamma)$, where $\afield{m}$ is the discrete image MVP, $\afield{m}'$ and $\nxhm'$ are corresponding test functions, and $H^{-1/2}(\curl; \Gamma)$ is a trace space~\cite{Monk_2003aa}.
	Here, the boundary condition~\eqref{eq:image_bc_strong} is weakly imposed by~\eqref{eq:image_bc_weak}, yielding a saddle-point problem~\cite{Brezzi_1991aa}. In this way, the quantity $\nxhm$, which will be needed for the calculation of $\afield{g}$, is already at hand.
	The weak sub-formulation~\eqref{eq:image_subform_weak} is eventually solved by a FE method employing standard edge shape functions. 
	\subsection{Reaction sub-formulation}\label{subsec:reaction}
	The reaction MVP $\Afield{g}$ is obtained by solving the BVP
	\begin{subequations}
		\begin{align}
			\nabla \times (\nu \nabla \times \Afield{g} ) &= \vec{J}_\mathrm{g}  & \text{in $V$}, \label{eq:adapted_pde_strong} \\
			\nvec \times \Afield{g} &= 0 & \text{on $\partial V$} \label{eq:adapted_bc_strong}
		\end{align}%
		\label{eq:adapted_subform_strong}%
	\end{subequations}%
	with $\vec{J}_\mathrm{g} = \vec{K}_\mathrm{g} \, \delta_\Gamma$, where $\vec{K}_\mathrm{g}$ is the surface current density~\eqref{eq:surface_current_density} and $\delta_\Gamma$ the delta distribution function defined by 
	\begin{equation}
		\int\limits_V f \, \delta_\Gamma \, \dd V = \int\limits_\Gamma f \, \dd S \quad \forall f.
	\end{equation}
	Here, $\Hfield{s} = \nu_0 \nabla \times \Afield{s}$ and $\Hfield{m} = \nu_0 \nabla \times \Afield{m}$ are the source and image magnetic field strengths, respectively.
	Visually, the current excitation in $\Va$ has been shifted onto the interface surface $\Gamma$. 
	The weak formulation of~\eqref{eq:adapted_subform_strong} reads:\\
	Find $\afield{g} \in H_0(\curl;V)$ s.t.
	\begin{equation}
		\begin{aligned}
			( \nu \nabla \times \afield{g}, \nabla \times \afield{g}')_V =& \\
			(\nvec \times \hfield{s},\afield{g}')_\Gamma &+ (\nxhm, \afield{g}')_\Gamma
		\end{aligned}
	 \label{eq:adapted_pde_weak}
	\end{equation}
	$\forall \afield{g}' \in H_0(\curl;V)$, where $\afield{g}'$ is a test function, and
	\begin{equation}
		H_0(\curl;V) = \lbrace \afield{} \in H(\curl;V)~:~ \gamma_{\partial V}(\Afield{}) = 0 \rbrace
	\end{equation}
	is chosen in order to fulfill~\eqref{eq:adapted_bc_strong}. The weak sub-formulation~\eqref{eq:adapted_pde_weak} is solved by a FE method employing standard edge shape functions. 
	Finally, the total MVP $\afield{}$ is composed of $\afield{g}$, $\afield{m}$ and $\afield{s}$ following~\eqref{eq:ansatz}. 
	
	\subsection{Treatment of nonlinearities}\label{subsec:nonlinear}
	Note that $\afield{g}$ corresponds to the full MVP in the domain $\Vi$. This facilitates the treatment of nonlinearities, as only $\afield{g}$ is affected by a nonlinear reluctivity $\nu = \nu(\rr, \vec{B}(\rr))$. To approximate the nonlinear material characteristic, a standard nonlinear iteration scheme can be employed such as fix-point iteration or Newton's method~\cite{Hauser_2009aa}, which is then only applied to the reaction sub-problem~\eqref{eq:adapted_pde_weak}. Other than that, no further modifications are needed to take nonlinearities in the updated RMVP formulation into account.

	\begin{figure}[t]
		\centering 
		\begin{tikzpicture}[scale=.7]
			\draw[fill=black!30!white] (0,0) circle (2);
			\draw[red!40!blue,fill=white, very thick] (0,0) circle (1.25);
			\fill[red!40!blue] (0.8, 0) circle (0.05);
			\node[red!40!blue] at (0.8, 0) [anchor=south] {$I$};
			\node[red!40!blue] at (270:1.25) [anchor=north] {$\Gamma$};
			\node at (110:0.65) {$\Va$};
			\node at (110:1.65) {$\Vi$};
		\end{tikzpicture}
		\caption{2D case study model: An infinitely long eccentric line current $I$ in an air domain $\Va$ surrounded by an infinitely long iron tube $\Vi$.}
		\label{fig:ecc_setup}
	\end{figure}
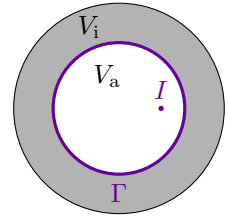
	
	\begin{figure}[t]
		\centering 
		\begin{subfigure}{1\columnwidth}
			\centering 
			\includegraphics[height=3cm]{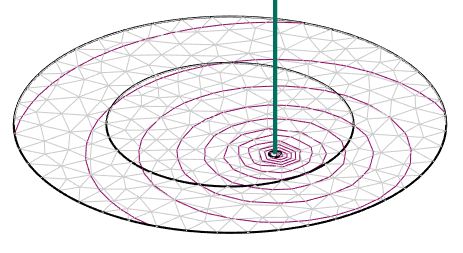}
			\caption{Equipotential lines of the source MVP $\vec{A}_\mathrm{s}$ in the eccentric wire case study calculated by~\eqref{eq:biot-savart-integral}. The source current is depicted as a green line. For the sake of visualization, $\vec{A}_\mathrm{s}$ is calculated in the whole domain instead of only at the interface boundary.}
			\label{fig:ecc_as}
		\end{subfigure}
		\begin{subfigure}{1\columnwidth}
			\centering 
			\includegraphics[height=3cm]{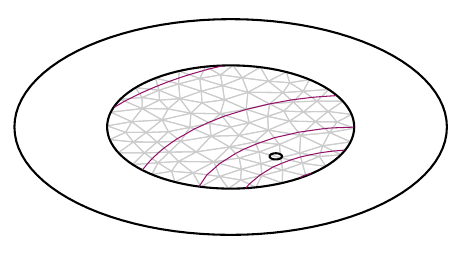}
			\caption{Equipotential lines of the image MVP $\vec{A}_\mathrm{m}$ in the eccentric wire case study calculated by~\eqref{eq:image_subform_weak}. The computation is only done in the sub-domain $\Va$.}
			\label{fig:ecc_am}
		\end{subfigure}
		\begin{subfigure}{1\columnwidth}
			\centering 
			\includegraphics[height=3cm]{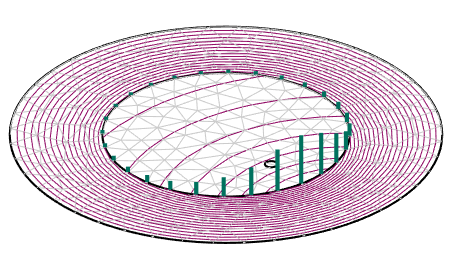}
			\caption{The green lines visualize the surface current density $\vec{J}_\mathrm{g}$ at $\Gamma$ determined by~\eqref{eq:surface_current_density}. The equipotential lines of the reaction MVP $\vec{A}_\mathrm{g}$ in the eccentric wire case study calculated by~\eqref{eq:adapted_pde_weak}.}
			\label{fig:ecc_ag}
		\end{subfigure}
		\begin{subfigure}{1\columnwidth}
			\centering 
			\vspace{0.25cm}
			\includegraphics[height=3cm]{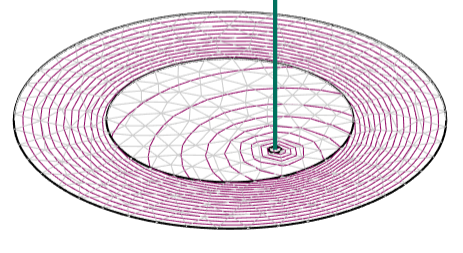}
			\caption{Equipotential lines of the total MVP $\vec{A} = \vec{A}_\mathrm{s}+\vec{A}_\mathrm{m}+\vec{A}_\mathrm{g}$ in the eccentric wire case study. The source current is depicted as a green line.}
			\label{fig:ecc_atot}
		\end{subfigure}
		\caption{The MVP components obtained in the sub-formulations of the RMVP formulation and the resulting total MVP for the case study of the eccentric wire in an iron ring.}
		\label{fig:ecc_components}
	\end{figure}
	
	\subsection{Illustration: Eccentric line current in an iron tube}
	Consider as a case study an infinitely long eccentric line current $I$ in air surrounded by an infinitely long iron tube. This model's 2D cross-section is shown in Fig.~\ref{fig:ecc_setup}. $\Va$ is defined as the air and coil region within the iron tube, while the iron tube itself is chosen as $\Vi$. Thus, the interface $\Gamma$ is the circular boundary between those two domains. 
	
	The RMVP formulation is applied to compute the MVP and magnetic flux density caused by the direct current $I$. Fig.~\ref{fig:ecc_as} shows the source MVP $\vec{A}_\mathrm{s}$, which is obtained by evaluating the Biot-Savart integral~\eqref{eq:biot-savart-integral}. It represents the MVP as if the eccentric line current were located in free space. The corresponding image MVP $\vec{A}_\mathrm{m}$ calculated by~\eqref{eq:image_subform_weak} is seen in Fig.~\ref{fig:ecc_am}. Fig.~\ref{fig:ecc_ag} visualizes the surface current density $\vec{K}_\mathrm{g}$ determined by~\eqref{eq:surface_current_density}, from which the reaction MVP $\Afield{g}$ calculated using~\eqref{eq:adapted_pde_weak} is shown as flux lines in the same figure.
	The superposition of these three sub-solutions leads to the total MVP $\vec{A}$ as shown in Fig.~\ref{fig:ecc_atot}.

	\section{Numerical Studies}\label{sec:studies}
	\subsection{Benchmark model: 2D racetrack coil}\label{subsec:2d_racetrack_coil}
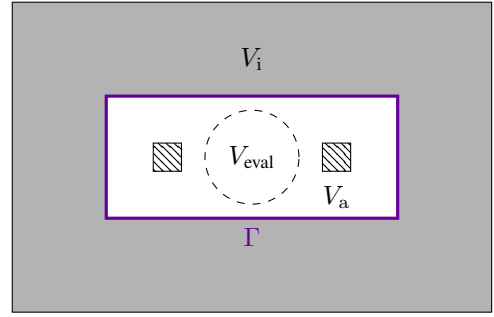
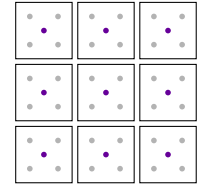
\begin{figure}[t]
		\centering 
		\begin{subfigure}{1\columnwidth}
			\centering 
			\begin{tikzpicture}[scale=.0125]
				\draw[fill=black!30!white] (-255, -165) rectangle (255, 165);
				\draw[fill=white, draw=red!40!blue, very thick] (-155, -65) rectangle (155, 65);
				\draw[dashed] (0, 0) circle (50);
				\draw[pattern=north west lines] (75, -15) rectangle (105, 15);
				\draw[pattern=north west lines] (-75, -15) rectangle (-105, 15);
				\node[red!40!blue] at (0, -65) [anchor=north] {$\Gamma$};
				\node at (0, 105) {$\Vi$};
				\node at (90, -42) {$\Va$};
				\node at (0, 0) {$V_\text{eval}$};
			\end{tikzpicture}
			\caption{Two coil windings (hatched rectangles, see Fig.~\ref{fig:rtc_detail} for a detailed sketch) in air (white, denoted as $\Va$) surrounded by an iron yoke (gray, denoted as $\Vi$). The interface boundary between iron and air is denoted as $\Gamma$. The dashed circular domain represents the evaluation domain for the magnetic energy.}
			\label{fig:rtc_total}
		\end{subfigure}
		\begin{subfigure}{1\columnwidth}
			\centering 
			\vspace{0.25cm}
			\begin{tikzpicture}[scale=.075]
				\draw (-5, -5) rectangle (5, 5);
				\fill[red!40!blue] (0, 0) circle (0.5cm);
				\fill[black!30!white] (-2.5, -2.5) circle (0.5cm);
				\fill[black!30!white] (-2.5, 2.5) circle (0.5cm);
				\fill[black!30!white] (2.5, -2.5) circle (0.5cm);
				\fill[black!30!white] (2.5, 2.5) circle (0.5cm);
				\draw (-6, -5) rectangle (-16, 5);
				\fill[red!40!blue] (-11, 0) circle (0.5cm);
				\fill[black!30!white] (-8.5, -2.5) circle (0.5cm);
				\fill[black!30!white] (-8.5, 2.5) circle (0.5cm);
				\fill[black!30!white] (-13.5, -2.5) circle (0.5cm);
				\fill[black!30!white] (-13.5, 2.5) circle (0.5cm);
				\draw (-17, -5) rectangle (-27, 5);
				\fill[red!40!blue] (-22, 0) circle (0.5cm);
				\fill[black!30!white] (-19.5, -2.5) circle (0.5cm);
				\fill[black!30!white] (-19.5, 2.5) circle (0.5cm);
				\fill[black!30!white] (-24.5, -2.5) circle (0.5cm);
				\fill[black!30!white] (-24.5, 2.5) circle (0.5cm);
				\draw (-5, 6) rectangle (5, 16);
				\fill[red!40!blue] (0, 11) circle (0.5cm);
				\fill[black!30!white] (-2.5, 8.5) circle (0.5cm);
				\fill[black!30!white] (-2.5, 13.5) circle (0.5cm);
				\fill[black!30!white] (2.5, 8.5) circle (0.5cm);
				\fill[black!30!white] (2.5, 13.5) circle (0.5cm);
				\draw (-6, 6) rectangle (-16, 16);
				\fill[red!40!blue] (-11, 11) circle (0.5cm);
				\fill[black!30!white] (-8.5, 8.5) circle (0.5cm);
				\fill[black!30!white] (-8.5, 13.5) circle (0.5cm);
				\fill[black!30!white] (-13.5, 8.5) circle (0.5cm);
				\fill[black!30!white] (-13.5, 13.5) circle (0.5cm);
				\draw (-17, 6) rectangle (-27, 16);
				\fill[red!40!blue] (-22, 11) circle (0.5cm);
				\fill[black!30!white] (-19.5, 8.5) circle (0.5cm);
				\fill[black!30!white] (-19.5, 13.5) circle (0.5cm);
				\fill[black!30!white] (-24.5, 8.5) circle (0.5cm);
				\fill[black!30!white] (-24.5, 13.5) circle (0.5cm);
				\draw (-5, 17) rectangle (5, 27);
				\fill[red!40!blue] (0, 22) circle (0.5cm);
				\fill[black!30!white] (-2.5, 19.5) circle (0.5cm);
				\fill[black!30!white] (-2.5, 24.5) circle (0.5cm);
				\fill[black!30!white] (2.5, 19.5) circle (0.5cm);
				\fill[black!30!white] (2.5, 24.5) circle (0.5cm);
				\draw (-6, 17) rectangle (-16, 27);
				\fill[red!40!blue] (-11, 22) circle (0.5cm);
				\fill[black!30!white] (-8.5, 19.5) circle (0.5cm);
				\fill[black!30!white] (-8.5, 24.5) circle (0.5cm);
				\fill[black!30!white] (-13.5, 19.5) circle (0.5cm);
				\fill[black!30!white] (-13.5, 24.5) circle (0.5cm);
				\draw (-17, 17) rectangle (-27, 27);
				\fill[red!40!blue] (-22, 22) circle (0.5cm);
				\fill[black!30!white] (-19.5, 19.5) circle (0.5cm);
				\fill[black!30!white] (-19.5, 24.5) circle (0.5cm);
				\fill[black!30!white] (-24.5, 19.5) circle (0.5cm);
				\fill[black!30!white] (-24.5, 24.5) circle (0.5cm);
			\end{tikzpicture}
			\caption{Detailed view of the coil winding configuration for $N_x = N_y = 3$. In the volumetric case~\eqref{eq:volumetric}, each winding is modeled as a rectangle with a given volumetric current density. In the RMVP case~\eqref{eq:adapted_pde_weak}, each half-turn is represented by a set of points (purple and gray dots) representing line currents in or out of plane, depending on the current orientation. In the simplest case, each half-turn is modeled by a single line current located in the cross-section center (purple dots).}
			\label{fig:rtc_detail}
		\end{subfigure}
		\caption{Racetrack coil model used for the purpose of numerical studies.}
	\end{figure}

	The RMVP formulation is implemented in the freely available open-source FE solver GetDP~\cite{Dular_1998aa} and employed to carry out a 2D linear magnetostatic simulation of a racetrack coil in an iron yoke. To justify the 2D approximation, it is assumed that the racetrack coil is very long compared to its cross-section diameter, such that the effects at the coil end windings can be neglected at a central cross-section of the model. Fig.~\ref{fig:rtc_total} shows the geometry, which consists of two winding groups (hatched rectangles) containing wires with rectangular cross-section and embedded in an air domain $\Va$ (white), which is surrounded by an iron yoke $\Vi$ (gray). $\Gamma$ is the interface between $\Va$ and $\Vi$. For the iron yoke, a linear permeability of $\mu_\mathrm{i} = 4000 \mu_0$ has been chosen.
	Since the line currents come together with singularities, the magnetic energy strives for infinity. Therefore, the magnetic energy considered below is evaluated in a sub-domain, which is the dashed circular domain $V_\text{eval}$ within $\Va$.
	
	The configuration of each winding group is shown in Fig.~\ref{fig:rtc_detail}. 
	In the 2D reference model, each half-turn of the coil is modeled as a surface (black rectangles in Fig.~\ref{fig:rtc_detail}) with a given surface current density. 
	In the 2D RMVP setting, each half-turn is discretized by a set of points (see Fig.~\ref{fig:rtc_detail}, gray and purple dots), which represent line currents in or out of plane, depending on the current orientation. In the simplest case, a half-turn is modeled by a single line current located in the cross-section center as seen in Fig.~\ref{fig:rtc_detail} (purple dots). 
	
	\begin{figure}[t]
		\centering 
		\includegraphics[width=.85\columnwidth]{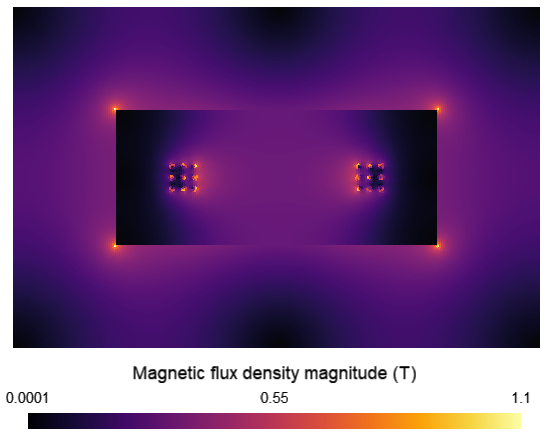}
		\caption{Magnetic flux density magnitude in the racetrack coil obtained by the updated RMVP formulation.}
		\label{fig:rtc_bfield}
	\end{figure}

	The results of the RMVP approach are verified against reference values obtained by a conventional 2D FE simulation modeling the half-turns as surfaces and solving~\eqref{eq:volumetric} while using a very fine mesh. The resulting magnetic flux density magnitude of the proposed method is shown in Fig.~\ref{fig:rtc_bfield}, where the maximal value has been capped to that of the reference solution. Particularly high fields are obtained at the four inner corners of the iron yoke due to the sharp geometry as well as the linear material properties and, of course, at the line currents due to their singular nature.
	
	\subsection{Convergence Analysis}
	
		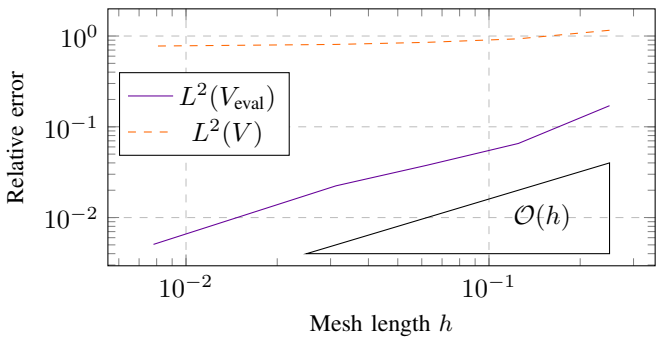
\begin{figure}[t]
		\centering 
		\begin{tikzpicture}
			\begin{loglogaxis}[
				height=5cm,
				width=1\columnwidth,
				ylabel={\small Relative error},
				xlabel={\small Mesh length $h$},
				y label style = {at={(axis description cs:0,0.5)}},
				legend style={at={(axis cs: 2.2e-2, 4e-1)}},
				ymajorgrids=true,
				xmajorgrids=true,
				grid style=dashed,
				]
				\addplot[color=red!40!blue] table [x={h_mesh}, y={rel_error_magnetic_energy_in_eval}, col sep=comma] {./img/linear_3x3_comparison.csv};
				%
				%
				\addplot[color=red!60!yellow, dashed] table [x={h_mesh}, y={rel_error_total_energy}, col sep=comma] {./img/linear_3x3_comparison.csv};
				\draw (axis cs: 0.025, 4e-3) -- (axis cs: 0.25, 4e-3) -- (axis cs: 0.25, 4e-2) --cycle;
				\node at (axis cs: 1.5e-1, 1e-2) {$\mathcal{O}(h)$};
				\legend{$L^2(V_\text{eval})$, $L^2(V)$};
			\end{loglogaxis}
		\end{tikzpicture}
		\caption{Linear convergence of the $L^2$-error in the evaluation domain (solid, purple) w.r.t.~the mesh length for the RMVP formulation. The $L^2$-error in the whole domain (dashed, orange) does not converge due to the singularities.}
		\label{fig:convergence}
	\end{figure}

	For the convergence analysis, the $L^2$-error w.r.t.~the volumetric reference solution is considered. The line currents have to be excluded from the considered domain, as otherwise the $L^2$-error would not converge due to the singularities. Therefore, a sub-domain $V_\text{eval} \subset V$ is chosen, such that $\mathcal{L}' \nsubseteq V_\text{eval}$ as seen in Fig.~\ref{fig:rtc_total} (circular region).
	
	For lowest order FEs, one would normally expect a quadratic convergence of the $L^2$-error. However, this only holds if, among other criteria, the right-hand side of the BVP is in $L^2(V)$~\cite{Ern_2018aa}. In the reaction sub-problem~\eqref{eq:adapted_subform_strong}, the right-hand side is a Dirac source term, making the problem not regular. Although classical convergence results are therefore invalid here, it has been shown that the $L^2$-error of 2D elliptic problems with Dirac source terms converges linearly~\cite{Scott_1973aa}. Indeed, this is observed for the $L^2(V_\text{eval})$-error in Fig.~\ref{fig:convergence} (solid purple line).
	
	Varying the size of $V_\text{eval}$, and thereby the remoteness of $\partial V_\text{eval}$ to the line currents, neither improves nor worsens the linear convergence behavior, as long as the $V_\text{eval}$ excludes the line currents or their immediate neighborhood. Then, the magnetic energy gets highly overestimated and does not converge at all (see Fig.~\ref{fig:convergence}, dashed orange line).

	\subsection{Performance comparison to the original formulation}
	The runtimes of the original and updated RMVP formulations are compared for the 2D magnetostatic linear simulation of the benchmark model with more than $100,000$ degrees of freedom on a standard workstation. Both the total runtimes and the runtimes of only the Biot-Savart integral evaluation are measured. Additionally, the updated RMVP simulation is carried out for the optimal case that only the source MVP on $\Gamma$ has to be computed as well as the worst case, in which the source MVP in $\Va$ is wanted. The latter would be the case if one is interested in the magnetic field in the whole aperture of a magnet. In practice, magnet designers are often interested in field values at particular points, e.g.~in a circular curve for a multipole coefficient analysis to investigate the field quality of the magnet~\cite{Russenschuck_2010aa}. Therefore, the average runtime is expected to range between those two extreme cases.
	
\begin{figure}[tb]
		\centering 
		\begin{tikzpicture}
			\begin{axis} [xbar = .05cm,
				bar width = 10pt,
				width=.8\columnwidth,
				height=5cm,
				xmin = 0, 
				xmax = 3600, 
				ymin = 0,
				ymax = 2,
				ytick={0, 1, 2},
				yticklabels={{\small Updated (on $\Gamma$)}, {\small Updated (on $\Va$)}, {\small Original (on $\Va \cup \Vi$)} },
				enlarge y limits = {abs = .8},
				enlarge x limits = {value = .25, upper},
				xlabel={\small Time (s)},
				]
				\addplot[fill=red!60!yellow] coordinates {(3563.628, 2) (461, 1) (7.341, 0)};
				\addplot[fill=red!40!blue] coordinates {(3581.875, 2) (467.162, 1) (12.489, 0)};
				\node at (axis cs: 3600, 2) [anchor=south west] {\small $3581\,$s};
				\node at (axis cs: 3600, 2) [anchor=north west] {\small $3563\,$s};
				\node at (axis cs: 470, 1) [anchor=south west] {\small $467\,$s};
				\node at (axis cs: 461, 1) [anchor=north west] {\small $461\,$s};
				\node at (axis cs: 12, 0) [anchor=south west] {\small $12\,$s};
				\node at (axis cs: 10, 0) [anchor=north west] {\small $7\,$s};
			\end{axis}
		\end{tikzpicture}
		\caption{Runtime comparison between the original and updated RMVP. Both the total times (purple) and the proportions of the Biot-Savart integral evaluations alone (orange) are depicted. For the updated RMVP case, the worst and best case with computation of $\vec{A}_\text{s}$ in $\Va$ and $\Gamma$ has been considered, respectively.}
		\label{fig:runtimes}
	\end{figure}
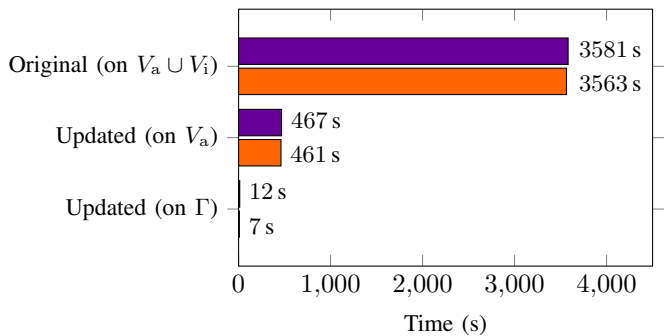

	Figure~\ref{fig:runtimes} shows the measured total runtimes (purple) and the portions of the Biot-Savart law evaluation therein (orange). The following two observations are made, from which two important conclusions are drawn:
	\begin{enumerate}
		\item The Biot-Savart integral computation heavily dominates the total runtime of both RMVP formulations. Therefore, improving that computation will significantly enhance the whole RMVP procedure. In favor of the RMVP formulation, the source field computation has been a research topic for decades and several efficient solution techniques exist to improve runtime, e.g., by utilizing a reduced scalar potential~\cite{Webb_1989aa, Biro_1993ab}, exploiting the existence of closed form expressions for conductors of specific geometric shapes~\cite{Gyimesi_1993aa}, or employing fast-multipole methods~\cite{Groh_2005aa}. Last but not least, one could also parallelize the computation of the Biot-Savart integrals.
		\item The updated RMVP formulation is by far computationally superior to the original formulation. This holds true even for the worst case. It is even more obvious when used for the calculation of the magnetic field at particular points, along certain curves or in small-sized regions in the magnet's aperture. Then, the updated method impressively outshines the original procedure.
	\end{enumerate}

	Hence, Fig.~\ref{fig:runtimes} illustrates the performance gain of the updated RMVP formulation with respect to the standard formulation and at the same time, parallelization of the Biot-Savart solver as a straightforward and promising measure for further improvement.
	
	\subsection{Distance of the line currents to the interface boundary}
	\begin{figure}[t]
		\centering 
		\begin{tikzpicture}
			\begin{loglogaxis}[
				height=4cm,
				width=1\columnwidth,
				ylabel={\small Relative $L^2$-error},
				xlabel={\small Relative distance of $\Gamma$ to line current},
				y label style = {at={(axis description cs:0,0.5)}},
				legend style={at={(axis cs: 2.2e-2, 4e-1)}},
				ymajorgrids=true,
				xmajorgrids=true,
				grid style=dashed,
				]
				\addplot[color=red!40!blue] table [x={relative_distance}, y={L2reldiff_Bfield}, col sep=comma] {./img/error_analysis_gamma.csv};
				\draw (axis cs: 1e-1, 5e-5) -- (axis cs: 1e-1, 5e-3) -- (axis cs: 3.15e-2, 5e-3) --cycle;
				\node at (axis cs: 7e-2, 1.5e-3) {$\mathcal{O}(\Delta^{-4})$};
				%
			\end{loglogaxis}
		\end{tikzpicture}
		\caption{Polynomial convergence of the relative $L^2$-error of the magnetic flux density w.r.t.~the relative distance of $\Gamma$ to the line current. The numerical quality is spoiled when the line current is located in the immediate neighborhood of $\Gamma$.}
		\label{fig:error_analysis_gamma}
	\end{figure}
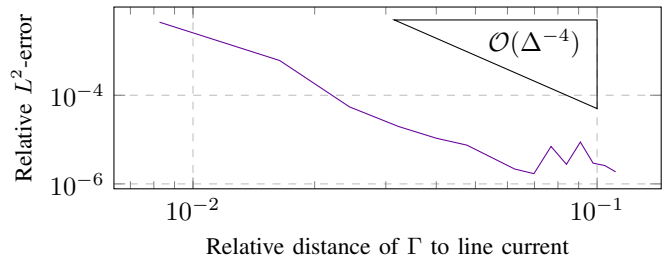
	The numerical quality of the updated RMVP formulation significantly depends on the distance of the line currents to $\Gamma$, which typically coincides with the interface to the iron yoke, but does not have to. To study that behavior, an eccentric line current in air with homogeneous Dirichlet boundary conditions is considered. For this problem, an analytical solution exists and will serve as a reference. While the position of the line current is fixed, the radius of $\Gamma$, $R_\Gamma$, which is an artificial rather than a material interface, is changed. Let $\delta$ denote the distance of the line current to $\Gamma$. The $L^2$-error of the magnetic flux density is then computed for each configuration and plotted against the relative distance of $\Gamma$ to the line current $\Delta = \nicefrac{\delta}{R_\Gamma}$ in Fig.~\ref{fig:error_analysis_gamma}. It is observed that the numerical quality highly worsens when the line current is positioned in the immediate neighborhood of $\Gamma$ ($\delta \rightarrow 10^{-4}\,$m or $\Delta \rightarrow 10^{2}$). However, the error rapidly decreases with increasing distance $\delta$: Already a tiny gap of $\delta \geq 10^{-3}\,$m or $\Delta \rightarrow 10^{-1}$ between $\Gamma$ and the line current is sufficient to eliminate this source of error completely. Since it is realistic for coils and current sources in general to be positioned a few millimeters away from the iron yoke, this numerical behavior is expected to be negligible in practical applications or at least easy to mitigate by substituting the outer coils by line currents which keep sufficient distance to $\Gamma$.

	\subsection{Application to three dimensions: 3D racetrack coil}
	The RMVP formulation as presented in Sec.~\ref{sec:rmvp} can be applied to a 3D model with only slight modifications compared to the 2D procedure:
	\begin{enumerate}
		\item The interface $\Gamma$ between $\Va$ and $\Vi$ is now represented by a 2D surface instead of a 1D curve.
		\item The coils are now being modeled by 1D current lines instead of 0D points. Accordingly, the general Biot-Savart expression~\eqref{eq:biot-savart-integral} has to be utilized. For the numerical integration, the current lines are approximated using discrete line elements.
		\item The MVP can now depend on all three space coordinates and can point in any direction, instead of depending only on $x$ and $y$ and featuring only a $z$-component as in the 2D approximation. Appropriately, 3D edge shape functions have to be used for the FE discretization of the 3D MVP.
		\item The MVP is not intrinsically gauged as it is the case for 2D problems~\cite{Russenschuck_2010aa}. Therefore, an explicit gauging has to be introduced in order to enforce a unique solution~\cite{Bossavit_1990aa}. Here, a Coulomb gauge~\cite{Russenschuck_2010aa} is employed, but one could also use a different gauging technique such as a tree-cotree gauge~\cite{Manges_1995aa}.
	\end{enumerate}
	Apart from those aspects, the core RMVP formulation does not change for 3D problems. 
	
	\begin{figure}[t]
		\centering 
		\includegraphics[width=.85\columnwidth]{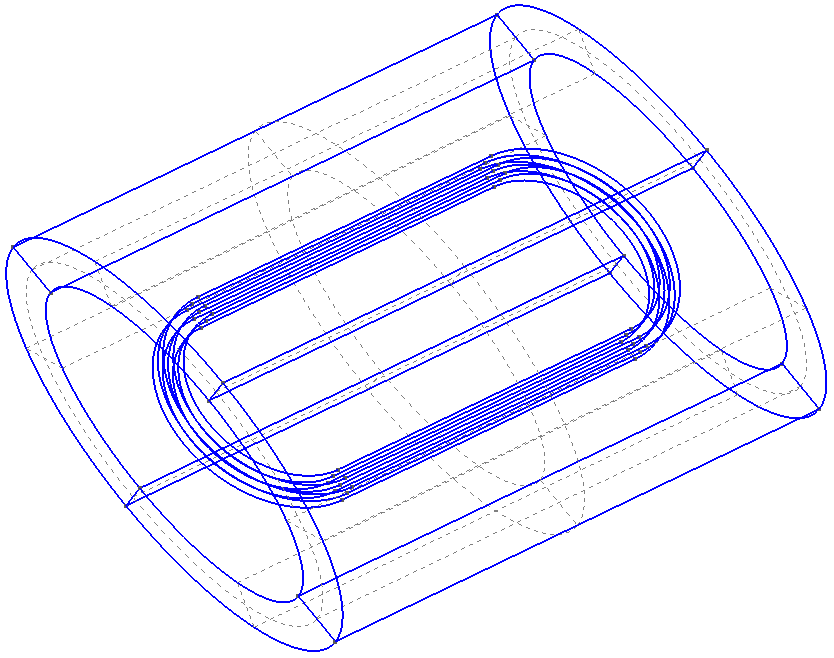}
		\caption{Geometry of the 3D racetrack coil model consisting of nine windings, which are represented by 1D closed curves and which are surrounded by an elliptically shaped iron yoke.}
		\label{fig:3d_racetrack_coil}
	\end{figure}

	\begin{figure}[t]
		\centering 
		\includegraphics[width=1\columnwidth]{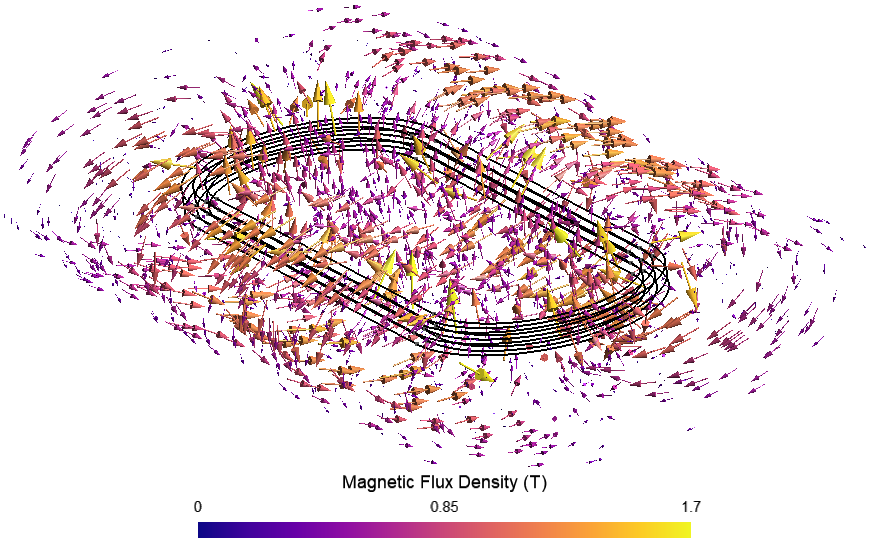}
		\caption{Magnetic flux density in the 3D racetrack coil model, computed by the updated RMVP formulation. The nine windings are modeled by 1D curve loops (black).}
		\label{fig:3d_rtc_field}
	\end{figure}

	\begin{figure}[t]
		\centering 
		\includegraphics[width=1\columnwidth]{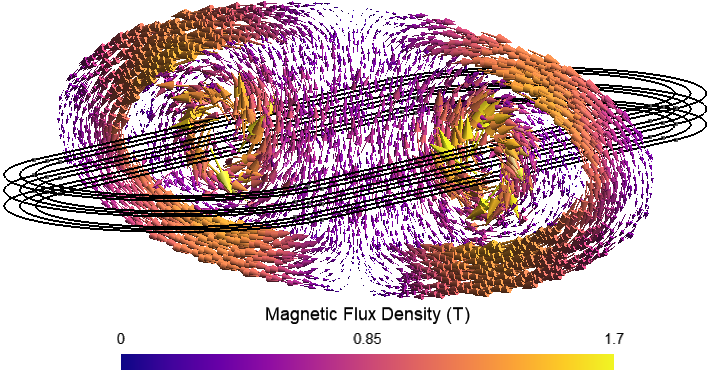}
		\caption{Magnetic flux density at the central $xy$-cross-section of the 3D racetrack coil, computed by the updated RMVP formulation. The nine windings are indicated in black.}
		\label{fig:3d_rtc_field_cs}
	\end{figure}
	
	The 3D RMVP formulation is demonstrated on a 3D racetrack coil model with an elliptical iron yoke and nine windings represented by nine 1D closed curves as shown in Fig.~\ref{fig:3d_racetrack_coil}. As for the 2D racetrack coil problem in Sec.~\ref{subsec:2d_racetrack_coil}, a linear iron permeability of $\mu_\mathrm{i} = 4000\mu_0$ is considered. 
	The resulting magnetic flux density is shown in Fig.~\ref{fig:3d_rtc_field} and Fig.~\ref{fig:3d_rtc_field_cs} in the whole model and on the central $xy$-cross-section, respectively. As in the 2D case, particularly high fields (recognizable as yellow arrows) are obtained near the line currents, where the field strives to infinity.

	\section{Simulation of the MQXA Quadrupole}\label{sec:quadrupole}
	
	\subsection{Setup in FiQuS/GetDP}
	\begin{figure}[t]
		\centering 
		\includegraphics[width=.9\columnwidth]{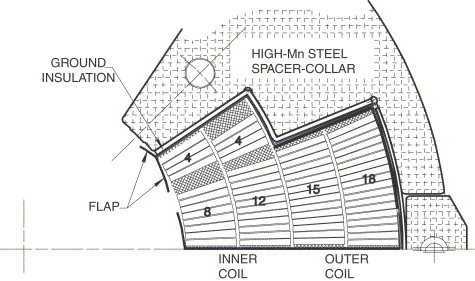}
		\caption{Coil concept drawing of one-eighth of the LHC MQXA quadrupole. Figure taken from~\cite{Ajima_2005aa}.}
		\label{fig:quadrupole_concept}
	\end{figure}
	The proposed RMVP formulation is employed to carry out a 2D magnetostatic nonlinear simulation of the LHC's MQXA low-beta quadrupole~\cite{Ostojic_2002aa, Ajima_2005aa} utilizing GetDP and FiQuS~\cite{Vitrano_2023aa}. 
	
	FiQuS is an open-source, free Python package, which enables a sustainable, consistent and reproducible workflow for computational engineers developing accelerator magnet models. This is especially important in the area of particle accelerator design, where the development of accelerator components spans over generations of researchers and engineers of multiple disciplines~\cite{Maciejewski_2023aa}. The updated RMVP formulation is implemented into FiQuS as a generalized template, which FiQuS adapts during runtime according to user-defined geometrical and physical information about the model.
	
	Fig.~\ref{fig:quadrupole_concept} shows one-eighth of the geometry of the considered MQXA quadrupole, which contains 61 windings (white rectangles), resulting into 488 windings in total~\cite{Ajima_2005aa}. Trapezoidal copper spacers (hatched trapezoids) help to achieve the desired coil form. An iron yoke (plaid area) surrounds the coil configuration.
	
	Each winding is approximated by one line current in its center, leading to 488 line currents and individual Biot-Savart integrals in the source sub-problem in total. A predefined BH-curve from the FiQuS material template database is used to model the iron yoke's nonlinear behavior. The nonlinear problem is solved by Newton's method, for which GetDP's built-in Jacobian functions are utilized.
	 
	\subsection{Simulation results} 
	\begin{figure}[t]
		\centering 
		\includegraphics[width=.85\columnwidth]{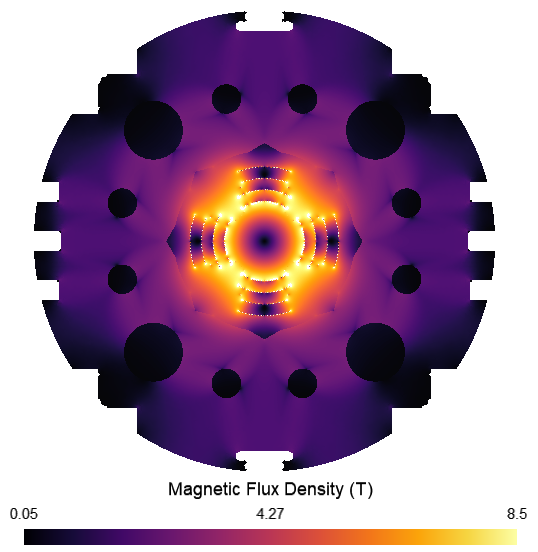}
		\caption{Magnetic flux density magnitude in the MQXA quadrupole computed by the RMVP formulation.}
		\label{fig:quadrupole_bfield}
	\end{figure}
	\begin{figure}[t]
		\centering 
		\includegraphics[width=.8\columnwidth]{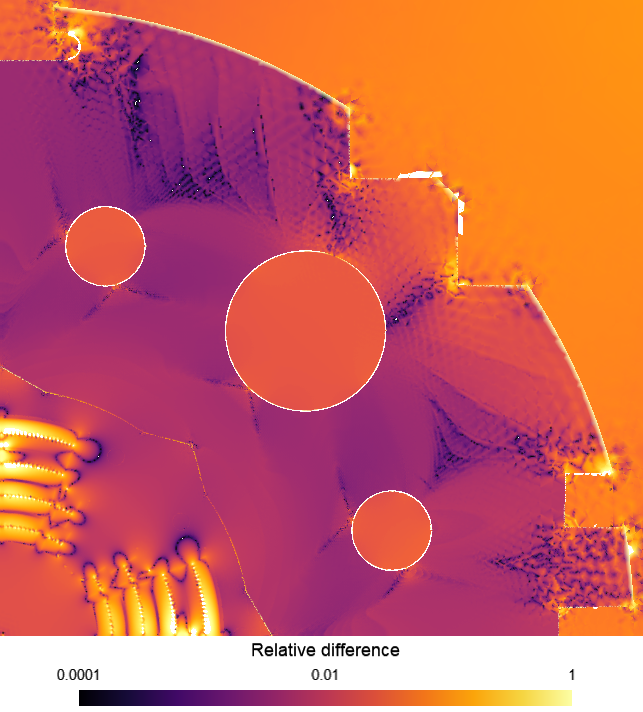}
		\caption{Relative difference between the magnetic flux density magnitude computed by the RMVP formulation and of the reference simulation performed by a conventional 2D FE solver.}
		\label{fig:relative_difference}
	\end{figure} 
	 
	Fig.~\ref{fig:quadrupole_bfield} shows the computed magnetic flux density in the magnet. The results are compared to a reference simulation performed by a conventional 2D FE method taking the windings into account by surface current densities. Fig.~\ref{fig:relative_difference} shows the relative difference $\epsilon_\mathrm{r}$ of the magnetic flux density magnitude obtained by the RMVP to that of the reference simulation. As expected, high discrepancies ($\epsilon_\mathrm{r} \geq 1$) occur in the neighborhood of the line currents, which represent singularities (clearly visible in Fig.~\ref{fig:relative_difference} as white dots). Outside the immediate neighborhood of the singularities, a good approximation is achieved with relative differences in the order of $\epsilon_\mathrm{r} = 10^{-3},\dots,10^{-1}$.
	An even better alignment between the RMVP and reference simulation is expected if the longish winding shapes were resolved by multiple line currents instead of just one at their centers.
	
	These results verify the procedure of treating nonlinear material characteristics as described in Sec.~\ref{subsec:nonlinear}. Furthermore, they demonstrate the operability of the updated RMVP implementation within FiQuS, with which the formulation could become accessible not only for CERN magnet designers, but also for the global magnet engineering community.

	\section{Conclusion}\label{sec:conclusion}
	This work proposed an updated RMVP formulation for accurate and fast magnetic field simulations of superconducting accelerator magnets. The formulation was postulated and verified against a 2D benchmark model. A runtime comparison showed that the proposed method clearly outperforms the original formulation. However, because of the Dirac source term occurring in one of the sub-problems, the $L^2$-error only shows a linear convergence for lowest order FEs instead of a quadratic one. This issue is well understood in the scientific computing community, and different techniques exist to improve the convergence order~\cite{Bertoluzza_2018aa}. Furthermore, the updated RMVP formulation was also employed to successfully simulate a 3D racetrack coil model.
	
	Finally, the updated RMVP procedure was embedded in the open-source and free Python package FiQuS, guaranteeing the greatest possible applicability in the magnet engineer community, and successfully employed to carry out a 2D nonlinear magnetostatic simulation of a quadrupole magnet. 
	
	The promising results of this work encourage an expansion of that method towards an updated simulation framework for HTS accelerator magnets, including physical formulations suitable for HTS magnets~\cite{Bortot_2020aa} and magnetization models for HTS tapes~\cite{VanNugteren_2016ab}. 
	
	
	\section*{Acknowledgment}
	\small{This work has been supported by CHART (\url{http://chart.ch}) in the context of the MagNum project, by the German BMBF project BMBF-05P18RDRB1, by the Graduate School Computational Engineering at TU Darmstadt, and by the DFG Research Training Group 2128 "Accelerator Science and Technology for Energy Recovering Linacs".}
	
	
	\printbibliography
	
\end{document}